\documentclass[prb,twocolumn,english,floatfix]{revtex4}
\usepackage{bm}
\usepackage{amsmath}
\usepackage{psfrag}
\usepackage{algorithmic}

\usepackage[pdftex]{graphicx}
\usepackage{epstopdf}

\makeatother
\begin{document}

\title{Modeling of solvent flow effects in enzyme catalysis under physiological conditions}

\author{Jeremy Schofield}
\email{jmschofi@chem.utoronto.ca}

\author{Paul Inder}
\email{pinder@chem.utoronto.ca}

\author{Raymond Kapral}
\email{rkapral@chem.utoronto.ca}

\affiliation{Chemical Physics Theory Group, Department of Chemistry, University
of Toronto, Toronto, ON M5S 3H6, Canada}

\date{\today}
\begin{abstract}%
\noindent
A stochastic model for the dynamics of enzymatic catalysis in explicit, effective solvents under physiological conditions is presented.  Analytically-computed first passage time densities of a diffusing particle in a spherical shell with absorbing boundaries are combined with densities obtained from explicit simulation to obtain the overall probability density for the total reaction cycle time of the enzymatic system. The method is used to investigate the catalytic transfer of a phosphoryl group in a phosphoglycerate kinase-ADP-bis phosphoglycerate system, one of the steps of glycolysis. The direct simulation of the enzyme-substrate binding and reaction is carried out using an elastic network model for the protein, and the solvent motions are described by multiparticle collision dynamics, which incorporates hydrodynamic flow effects. Systems where solvent-enzyme coupling occurs through explicit intermolecular interactions, as well as systems where this coupling is taken into account by including the protein and substrate in the multiparticle collision step, are investigated and compared with simulations where hydrodynamic coupling is absent. It is demonstrated that the flow of solvent particles around the enzyme facilitates the large-scale hinge motion of the enzyme with bound substrates, and has a significant impact on the shape of the probability densities and average time scales of substrate binding for substrates near the enzyme, the closure of the enzyme after binding, and the overall time of completion of the cycle.
\end{abstract}

\maketitle
\section{Introduction}

Biochemical reactions in the cell are often carried out through complex chemical networks consisting of many coupled elementary component steps~\cite{MBC94}. Even the elucidation of the molecular-level mechanism which underlies the operation of a single component in such networks is often a difficult task. Computer simulation is playing an increasingly important role in such mechanistic studies but direct simulation of many biochemical processes is challenging because they occur on a diverse range of scales. This fact has prompted the development of coarse-grain or mesoscopic methods that allow one to circumvent some of the difficulties related to dynamics that take place on long space and time scales~\cite{BR05,V08}. In enzyme kinetics long times scales can arise from the diffusive approach of the substrate to the enzyme and the conformational changes in the enzyme in the course of the catalytic reactions it carries out. There have been numerous simulation studies of the effects of diffusion on enzyme kinetics.~\cite{park08,zhou96,sung05,popov01,kim99}.  In this paper we describe how one may construct a mesoscopic model of an enzymatic cycle that incorporates the diffusive approach of substrates to the enzyme based on the solution of the diffusion equation, along with a particle-based description of the enzymatic reaction that involves protein conformational changes, release of the product and the return of the protein to its original conformation.

\begin{figure}[htbp]
\centerline{%
  \includegraphics[width=0.75\columnwidth]{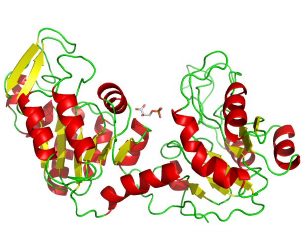}}
  \caption{The open conformation of phosphoglycerate kinase showing the N- and C-terminal domains of the protein.}
  \label{fig:pgk}
\end{figure}

The method is used to investigate a specific enzymatic reaction, 
\begin{equation}
bPG +ADP \mathrel{\mathop{\kern0pt {\rightleftharpoons}}\limits^{{PGK}}_{}} PG +ATP,
\end{equation}
catalyzed by the enzyme phosphoglycerate kinase (PGK). This reaction is an important step in the glycolysis network. In particular, we focus on the forward reaction that involves the transfer of a phosphoryl group from 1,3-bisphosphoglycerate (bPG) to ADP by the PGK enzyme to form 3-phosphoglycerate (PG) and ATP. (Often it is the reverse reaction that is studied experimentally due to the instability of bPG.~\cite{schmidt95}) Phosphoglycerate kinase is a monomeric protein of moderate size (416 amino acid residues in the human isozyme studied here) found in all living organisms, with a highly conserved amino acid sequence across different life forms. Its structure, consisting of two equal-sized domains labeled by the N- and C-termini of the protein, is well-adapted to selectively bind two substrates: bPG binds to the N-terminal, while the nucleotide substrates, MgATP or MgADP, bind to the C-terminal domain of the enzyme.  Structurally, the N- and C-domains consist of a 6-stranded parallel beta-sheet surrounded by alpha helices (see Fig.~\ref{fig:pgk}).

The mechanism for the enzymatic reaction, which involves large hinge bending motions of the domains of the protein~\cite{banks79,bernstein97,bernstein98}, has been the subject of many kinetic studies~\cite{schmidt95,geerlof97,geerlof05,varga06}. 
The activity of the enzyme requires both substrates to be bound.~\cite{bernstein97,varga09,auerbach97}  When both substrates bind, the enzyme undergoes a large-scale hinge-bending conformational change that brings the substrates close to one another to catalyze the dephosphorylation of bPG. In this ``closed'' conformation, the transition state is stabilized, lowering the free energy barrier for the transfer of a phosphoryl group. Upon transfer, the enzyme is forced into an open configuration and the PG and ATP products are released.

We shall be concerned with the enzymatic activity of PGK under physiological conditions in the cell where the binding process is diffusion limited.~\cite{footnote-phys-cond} The  binding process is well suited to be modeled as a two-step process in which first the substrates diffuse freely into a region near the enzyme, and then are drawn into the binding sites on the enzyme. Thus, it is reasonable to utilize a hybrid, stochastic procedure that combines analytical calculations with explicit simulation. The first step in the process of computing the distribution of time scales of the catalytic activity of the enzyme can be estimated by calculating first-passage times for the substrates moving into the vicinity of the enzyme, while the second step requires a more detailed dynamical simulation due to the influence of the enzyme on the dynamics of the substrate. There have been simulations of the domain motions of PGK using a variety of methods.~\cite{guilbert96,vaidehi00,blog07,palmai09,inoue10} Given the large size of the protein and the long time scales of the motions, a full molecular dynamics simulation of the second step, which involves binding of the substrates to the enzyme in solution, the hinge-bending motion of the enzyme-substrate complex, followed by the reaction of the substrates and final release of products coupled with the re-opening of the enzyme, is computationally demanding. Consequently, we develop a coarse grain description of this part of the enzymatic cycle that is particle-based, includes enzyme, substrates and solvent molecules explicitly and retains many features of full molecular dynamics.

The outline of the paper is as follows. The two steps of the enzymatic reaction dynamics, diffusive approach of enzyme and substrate and substrate binding and reaction, are described first. The mesoscopic model for the protein, substrates and solvent, along with a description of the interaction potentials that control the binding of the bPG substrate to the active site and conformational changes in the PGK protein, are the topics of Sec. ~\ref{sec:meso-protein}. Section~\ref{sec:diffusion} discusses the various time scales involved in the diffusive encounter between the substrates and the enzyme and shows how the relevant first-passage times can be computed analytically. The results of simulations of the dynamics are reported in Sec.~\ref{sec:sim} while the conclusions of the study are summarized in Sec.~\ref{sec:conc}.

\section{Protein and its catalytic activity in solution} \label{sec:meso-protein}

We consider a system containing PGK enzymes, along with substrate and solvent molecules. The enzyme exists in open and closed forms and binding of both substrates is necessary for large-scale conformational changes to occur.~\cite{bernstein98,auerbach97,varga06} We suppose that the ADP substrate is bound to the enzyme and construct a coarse-grain model of the protein interacting with the bPG substrate in the presence of solvent. As discussed below, under physiological conditions, ADP binds quickly and the rate of the enzymatic reaction is determined by the binding of bPG. The model of the enzymatic activity of PGK entails a description of the interactions of bPG with the enzyme as it binds to the active site, the conformational changes in the protein that lead to the reactive event and the release of product and return of the protein to its original conformation.

\subsection{Network model of PGK and interactions with substrate}

A coarse-grain network model of the PGK protein is constructed by replacing each amino acid residue with a single monomer bead and connecting the beads by links or bonds.~\cite{tirion96,maragakis05,tozzini05,V08} The bound ADP substrate is treated as one of the protein beads, while the bPG substrate is also described in a coarse-grained fashion as a single bead. The set of bead coordinates ${\bf R}^{N_P}=({\bf R}_1, {\bf R}_2,\dots, {\bf R}_{N_P})$ specifies the configuration of the protein (P) and we let ${\bf R}$ denote the coordinate of bPG, henceforth called the substrate (S). The construction of the potential energy function that is responsible for the protein conformational state and interactions between the protein and substrate are described in detail in Appendix A. Here we simply sketch the main elements that enter in the design of the potential function, $V_{PS}({\bf R}^{N_P},{\bf R}; \xi)$, that is able to describe both conformational states of the protein, the binding of bPG to the active site, and the resulting changes of protein conformational states that occur on substrate binding and product release~\cite{footnote-ake}.

To construct a network model for PGK, protein database configurations built from crystallographic data were analyzed to determine a set of pairwise interactions between residues.  Each of the 416 residues was represented by a single monomer bead in a linear polymer representation of the protein, with the position of each bead taken to be the Cartesian coordinates of the alpha-carbon of the peptide.  Both open and closed forms of the PGK molecule were taken from the initial and final protein database configurations generated from the morphing analysis of the conformational change between open and closed conformations~\cite{pgk1,pgk2} in the Database of Macromolecular Movements~\cite{proteinMorph}.

Pairs of beads separated by a distance $r < 10$  $\AA$ were recorded, generating separate lists of indices for open and closed conformations.  The interaction lists for open and closed configurations were then compared, and a set ${\mathcal B}_c$ of common interaction pairs or links were identified and assigned bond potentials in the following way. For links in ${\mathcal B}_c$ the bond length as well as the magnitude of the difference $r_{co}$ between the bond lengths in the open and closed conformations were computed. The links in ${\mathcal B}_c$ were then grouped into two new subsets, ${\mathcal B}_{hc}$ and ${\mathcal B}_{sc}$, containing hard (hc) or soft common (sc) links, respectively, based on the value of the separation distance $r_{co}$, where links with $r_{co}<4$ were identified as hard links.  The list of common links was then compared with the lists of open links and closed links. This process yielded $2891$ hard-common links and $519$ soft-common links. In this study, the ADP substrate is treated as a single bead that forms hard links with three different beads in the enzyme.

Pairs that exist in either the list of open links or the list of closed kinks but not in both were sorted into soft-open, ${\mathcal B}_{so}$, and soft-closed, ${\mathcal B}_{sx}$ sets, respectively. There are $448$ soft-open links (so), and $619$ soft-closed links (sx).

\begin{figure}[htbp]
\centerline{%
  \includegraphics[width=0.5\columnwidth]{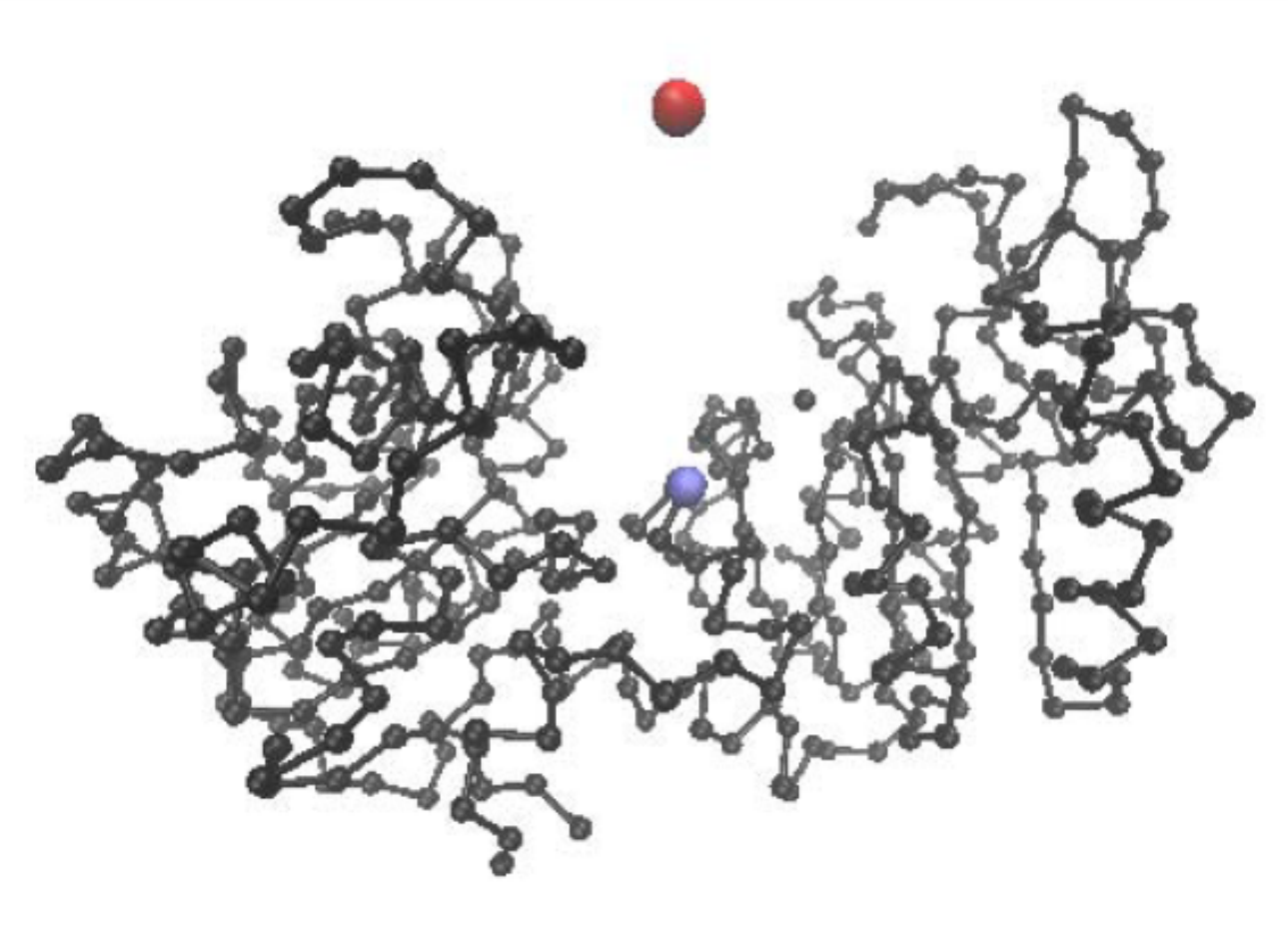}
  \includegraphics[width=0.5\columnwidth]{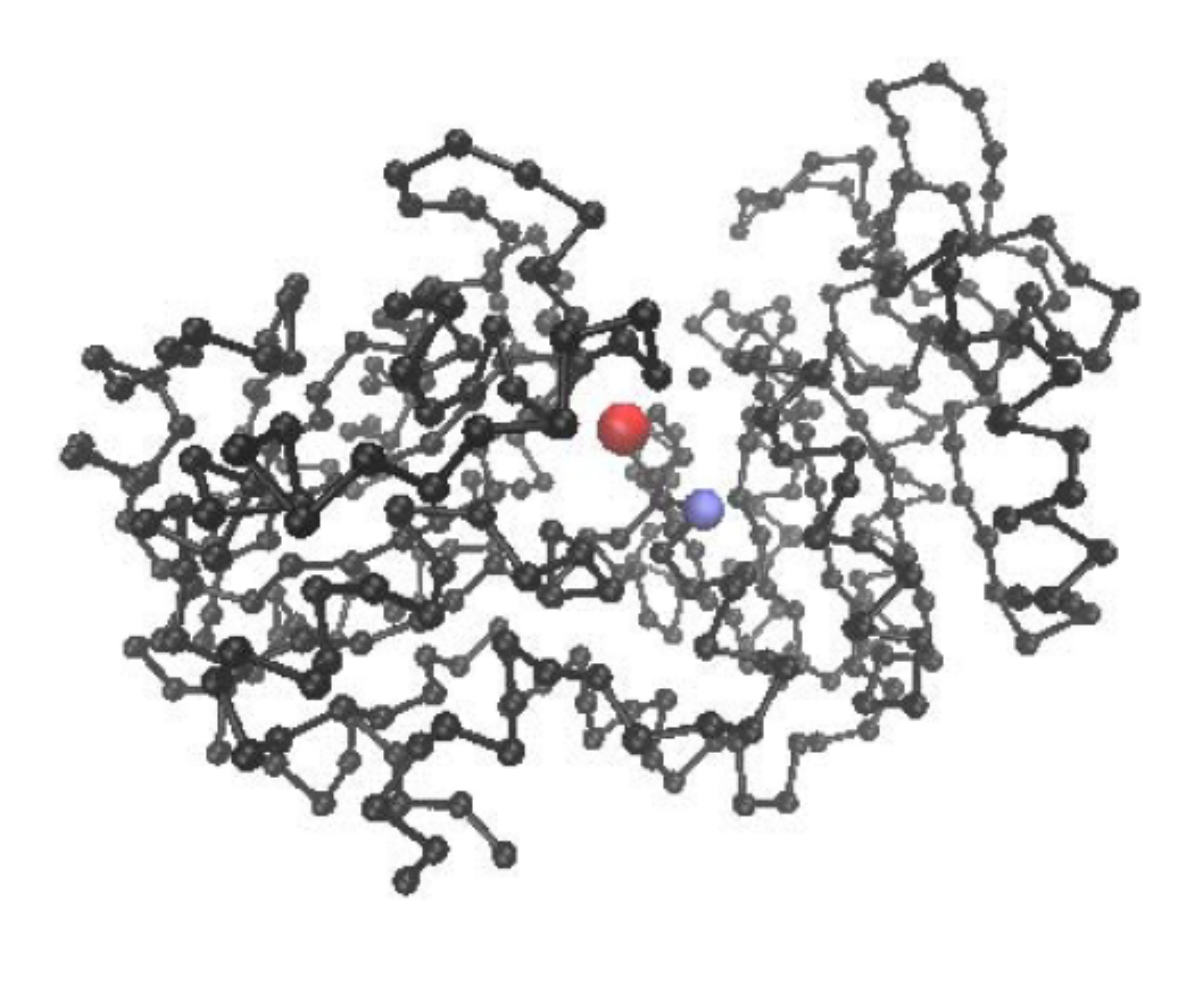}}
  \caption{(left) Open conformation of the network model of PGK showing the approach of bPG to the binding pocket of the enzyme. (right) Protein conformation after substrate binding has resulted in hinge closing to form the closed conformation.}
  \label{fig:op-cl-pgk}
\end{figure}

Before the enzymatic reaction can occur, bPG must bind to the active site of the enzyme. The binding pocket of the enzyme for this substrate was defined by beads with coordinates $({\bf R}_0^a, {\bf R}_1^a, {\bf R}_2^a)$, where $\bm{R}_0$, $\bm{R}_1$ and $\bm{R}_2$ are the coordinates of the alpha-carbon of the glycine residues $386$, $387$ and $388$ in the amino acid sequence of the PGK enzyme.  The binding interaction between the bPG substrate at position ${\bf R}$ and the enzyme was assumed to depend on both the distance between the substrate and the bead in the active site with coordinate ${\bf R}_1^a$, $|\bm{R} - \bm{R}_1^a| = R_{S1} $, as well as the orientation of the substrate with respect to a coordinate frame determined by three beads defining the binding pocket of the enzyme. As the substrate binds it triggers conformational changes in the protein that lead to hinge closing to bring the bPG and ADP substrates into proximity for the phosphoryl group transfer.  Consequently as bPG interacts with the protein in the course of binding to the active site, the open protein configuration is destabilized with respect to the closed configurations, driving the enzyme towards the closed conformation. To achieve this conformational change, the interaction potentials for the soft, non-common set of links were taken to depend on a reaction coordinate $\xi(R_{S1})$, which is a function of the distance between bPG and the active site.
The net effect of the combination of these contributions is a protein-substrate interaction potential, $V_{PS}({\bf R}^{N_P},{\bf R}; \xi)$, which can draw in the bPG substrate, bind it to the active site of the enzyme in the open configuration, and then cause the enzyme to undergo a conformational change from an open to closed configuration. The network model of the protein and the binding of the substrate to the open conformation leading to hinge closing is shown in Fig.~\ref{fig:op-cl-pgk}.
After binding has taken place, the phosphoryl group  transfer reaction is carried out by treating the reaction coordinate $\xi$ as an external control parameter whose value is determined probabilistically.  When the reaction is complete the closed configuration is unstable and the enzyme reopens, completing the cycle.

\subsection{Solvent and its interactions with the protein and substrate}

The system also contains $N_s$ solvent molecules with positions, ${\bm r}^{N_s}=({\bm r}_1,{\bm r}_2, \dots, {\bm r}_{N_s})$ and velocities, ${\bm v}^{N_s}=({\bm v}_1,{\bm v}_2, \dots, {\bm v}_{N_s})$. The solvent evolution is modeled by multiparticle collision (MPC) dynamics.~\cite{MalevanetsKapral99} In MPC dynamics there are no intermolecular potentials among solvent molecules. Instead, solvent molecules propagate in the absence of solvent-solvent interactions and undergo multiparticle collisions at discrete times $\tau$ that account for the effects of many real collisions during this time interval. More specifically, after the streaming step, solvent particles are assigned to cells with length $\ell$ for the purposes of carrying out multiparticle collisions.  The center-of-mass velocity ${\bm{v}}_c$ of particles in a cell is computed for each cell $c$, and the velocities of the solvent particles relative to the center-of-mass velocity are rotated around a randomly chosen axis by an angle chosen from a set of possible rotations.  This ``collision'' step conserves linear momentum, energy and particle number, and is consistent with hydrodynamic flow~\cite{Kapral08,GIKW09}.  The collision step for a particle $i$ in cell $c$ is therefore:
\begin{eqnarray}
\bm{v}_i^\prime = {\bm{v}}_c + \bm{\omega} \cdot (\bm{v}_i - {\bm{v}}_c),
\end{eqnarray}
where $\bm{v}_i^\prime$ is the post-collision velocity of particle $i$ and $\bm{\omega}$ is a rotation matrix.

When the system contains proteins and substrates dissolved in the solvent, the evolution is described by hybrid molecular dynamics-multiparticle collision (MD-MPC) dynamics.~\cite{MalevanetsKapral00} In such hybrid dynamics, while the solvent molecules interact among themselves through multiparticle collisions, they interact with the solute molecules through solvent-bead intermolecular forces, $V_{sb}$. The total potential energy of the system is therefore given by $V_T=V_{PS}+V_{sb}$ and Newton's equations of motion are used to evolve the system under this potential energy for time intervals $\tau$ between MPC events. This hybrid dynamics also satisfies the conservation laws and correctly describes hydrodynamic interactions among solute species and fluid flows in the solvent.

\subsubsection*{Penetrating solvent model}
Hydrodynamic interactions and solvent dissipation can also be included in a heuristic way by dropping the direct interactions between solvent and solute molecules and instead including the solute beads into the MPC step of the dynamics.~\cite{MY2000}  In this scheme, the solvent particles evolve freely between collision steps while the coordinates and momenta of the enzyme and substrate are evolved through Newton's equations of motion under the $V_{PS}$ potential function.

More specifically, to allow for interaction between the solvent and beads, the collision rule is modified to include the velocity of the beads in the local center-of-mass velocity of particles in a cell.  The center-of-mass velocity is computed for a cell $c$ containing $N_c$ solvent particles of equal mass $m$ and a single bead of mass $M$ and velocity $\bm{V}$ via:
\begin{equation}
\bm{v}_c = \frac{M}{M_T} \bm{V} + \frac{N_c m}{M_T}
\sum_{i=1}^{n_c} \bm{v}_i,
\end{equation}
where $M_T = N_c m + M$ is the total mass of particles in the cell and
$\bm{v}_i$ is the velocity of solvent particle $i$ in cell $c$.  The
collision rule for the penetrating solvent model with hydrodynamics is
defined as
\begin{eqnarray}
\bm{V}^\prime &=& \bm{v}_c + \bm{\omega} \cdot \left( \bm{V} -
  \bm{v}_c \right)\nonumber \\
\bm{v}_i^\prime &=& \bm{v}_c + \bm{\omega} \cdot \left( \bm{v}_i -
  \bm{v}_c \right),
\end{eqnarray}
for the bead velocity $\bm{V}$ and the solvent velocities $\bm{v}_i$.
Since the magnitude and direction are conserved in the rotation,
particle number, linear momentum and energy are globally conserved,
resulting in proper hydrodynamic flow.

\subsubsection*{Penetrating solvent without hydrodynamics}
For the purpose of assessing the importance of hydrodynamic interactions, it is useful to construct an alternative model in which the hydrodynamic effects are not present.~\cite{kikuchi02,kikuchi03,ripoll07,GIKW09} The collision rule for the penetrating solvent model can be modified by defining the center-of-mass velocity of particles in a cell to be
\begin{equation}
\bm{v}_c = \frac{M}{M_T} \bm{V} + \frac{N_s m}{M_T} \bm{v}_s,
\end{equation}
where $N_s$ is drawn from a Poisson distribution with mean value
$\rho V_c$, where $\rho$ is the number density of solvent in the system and $V_c$ is the cell volume. The total mass is $M_T = M + N_s m$, and $\bm{v}_s$ is an effective solvent velocity drawn from a Maxwell-Boltzmann distribution with mass $N_s m$.  Since this velocity is drawn at each collision step, the velocity of the solvent is uncorrelated from one collision step to another.  In this model, explicit solvent particle dynamics is replaced by the action of the collision operator. Since the velocity of the fluid is completely decorrelated after a single collision step, any dynamic correlations associated with a small value of the ratio of the mean free path to cell length strictly vanish.

\section{Enzymatic cycle dynamics}\label{sec:diffusion}

Complete enzymatic cycles can be simulated using the mesoscopic dynamical scheme described in the previous section. When the protein, substrates and solvent molecules are modeled as structureless particles, full MD-MPC dynamics has been used to study the effects of diffusion on enzyme kinetics.~\cite{chen11} However, in the conditions that pertain to the interior of a cell, even this multi-scale method will not be computationally efficient if both the internal dynamics of the enzymes and the diffusive motion are considered. Under physiological conditions the concentrations of both substrates in the cytoplasm are relatively small~\cite{garrett04,minakami65} (0.14 mM for ADP and 0.001 mM for bPG), while the enzyme concentration is roughly 0.1 mM. If the substrates and enzyme are uniformly distributed in the volume, the radius of the spherical volume around the enzyme containing a single substrate molecule is roughly $r_{{\rm ADP}}=142$~\AA~for ADP and $r_{{\rm bPG}}=734$~\AA~for bPG.  The sphere containing a single enzyme has a radius of $r_{{\rm PGK}}=158$ \AA. Estimating the viscosity of the cytoplasm to be roughly 5 times that of water, namely $\eta = 0.005$ Kg/(m-s), and assuming the substrates have an effective radius $R_S \approx 5$ \AA, the Stokes-Einstein law $D = k_BT/(6\pi \eta R_S)$ gives a value of $D=910$
$\mathrm{\AA}^2/\mu s$.  Given these conditions, we shall see that the ADP substrate binds typically before $5$~$\mu s$, whereas the binding time of the bPG is very broadly distributed over many decades and is the main factor determining the reaction time. For this reason we suppose that ADP is bound to the enzyme and focus on the binding of bPG.

From these considerations it is evident that the enzymatic dynamics has a significant diffusion-influenced component; therefore, it is computationally inefficient to follow individual trajectories of the diffusive dynamics of substrates and enzymes in the solvent for the long times needed for enzyme-substrate encounters. Consequently, it is useful to decompose the process into portions where the substrates diffuse in the solvent without directly interacting with proteins, and portions where these species interact through direct intermolecular forces. The diffusive portions of the dynamics can be treated to a good approximation by analytical methods, while in the interacting portions the mesoscopic dynamical scheme can be used to describe details of the binding, conformational changes and reaction. These considerations suggest a stochastic model for the cycle dynamics that combines these types of dynamical evolution.

\subsection{Stochastic model for enzyme dynamics}

\begin{figure}
\centerline{%
  \includegraphics[width=0.7\columnwidth]{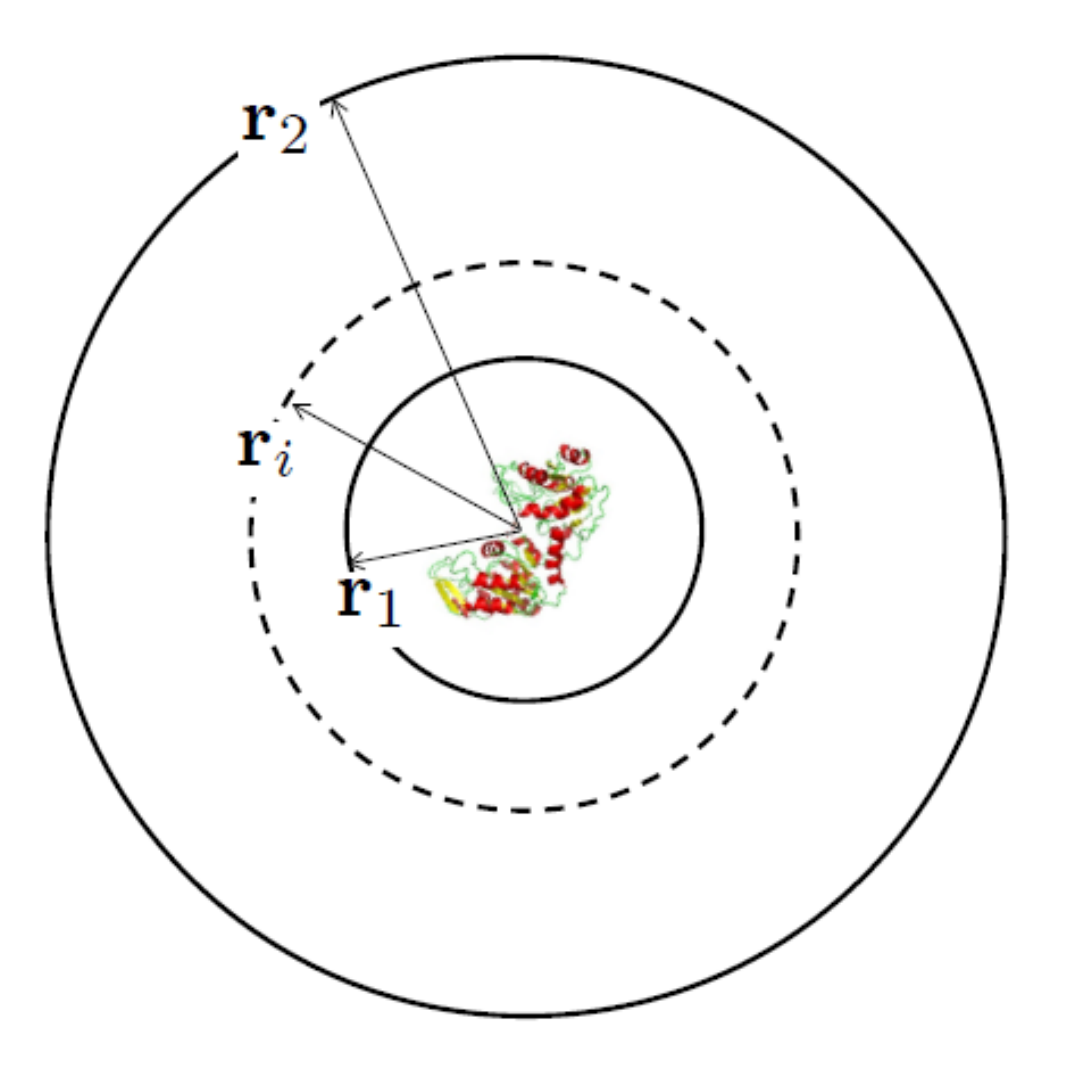}}
  \caption{Structure of the model. For the system considered here, we have chosen $r_2 = 31.6$, $r_i = 9$, and $r_1 =
7$ in simulation cell length units.  This choice of radial distances allows one to minimize the amount of numerical simulation required while allowing for
good statistics for various numerically computed densities.}
  \label{fig:onion}
\end{figure}

Initially, suppose the bPG substrate moves diffusively in a volume with radius $r_{{\rm bPG}}$ surrounding the enzyme without any influence on its motion due to the presence of an enzyme.  Since the concentration of enzyme is a factor of $100$ times that of the bPG, the number of enzymes in this volume should be Poisson distributed with an average number of $100$ enzymes in the volume if there is no correlation in the density of enzymes. We assume the binding of the bPG to any enzyme in the volume occurs in the following way:  At any given time, the bPG is within the spherical volume with radius $r_2$ of some enzyme, which is smaller than the volume around the enzyme that contains a single substrate molecule (see Fig.~\ref{fig:onion}).  The substrate can either diffuse to the binding region of this enzyme, or out of its volume.  The binding probability is dependent on how far the substrate is from the enzyme.  If the substrate diffuses out of the volume of the enzyme, the first passage time out of the spherical volume can be recorded.  Subsequently, the position of the bPG relative to another enzyme is assumed to be randomly distributed in the volume of this other enzyme, and the process is repeated until the substrate passes through the inner spherical volume of radius $r_1$ around an enzyme.

The point where the substrate passes through the inner sphere is uniformly distributed on the surface of the sphere. After passing through the inner sphere, the substrate will either bind to the enzyme or move out of the inner sphere and pass through a sphere of intermediate size (with radius $r_i$ with $r_1 \leq r_i \ll r_2$).  Since the dynamics
of the substrate is influenced by the presence of the enzyme and the solvent flow around it, the dynamics is no longer diffusive and must be simulated explicitly as described in the previous section. Starting from a uniformly chosen point on the surface of the sphere with radius $r_1$, if the substrate does not bind to the active site, the particle continues to diffuse starting from a radial distance of $r_i$ and either will be reabsorbed by the inner sphere or pass out of the volume through the outer sphere.

For most of the dynamical evolution, the substrate diffuses freely without explicit solvent flow effects or influence from the enzyme.
For this type of dynamics, analytical solutions to the diffusion equation can be used.  The final regime to be described consists of the dynamics of the substrate from the surface of the inner sphere with radius $r_1$ to the active site on the enzyme in the presence of solvent. This final regime should be simulated directly, since the hydrodynamic motion of the solvent influences both substrate and enzyme motion.

More specifically, the algorithm can be stated as follows:

(1) At the initial time if the substrate is at ${\bf r}_2$ a position ${\bf r}$, ${\bf r}_1 <{\bf r} <{\bf r}_2$, is randomly selected.

(2) Given a uniformly distributed random number $\xi_r \in [0,1]$, if $\xi_r \le P_1(r)$ the substrate is absorbed at the $r_1$ boundary, otherwise it is absorbed by the $r_2$ boundary. Here $P_1(r)$ is the probability that the substrate is absorbed at the $r_1$ boundary in the infinite time limit.

(3) If it is absorbed at $r_2$, a time is drawn from $P_2(t|r)$, the first-passage time density for absorption onto a sphere with radius $r_2$ starting a distance $r$ from center, and used to update the cycle time.

(4) If it is absorbed at $r_1$, a time is drawn from $P_1(t|r)$, the first-passage time density for absorption onto a sphere with radius $r_1$ starting a distance $r$ from center, and used to update the cycle time. Starting at $r_1$, a full mesoscopic dynamical simulation is then carried out until reaction occurs or the substrate reaches the $r_i$ boundary. If the dynamics results in a reaction, the time for this to occur is added to the cycle time and the enzymatic cycle is complete. If instead the substrate reaches $r_i$ without reaction, this time is added to the cycle and we return to step (2) to continue the dynamics until the cycle is complete. The boundary at $r_i$ is chosen to be significantly larger than $r_1$ to minimize the blocking effect of the enzyme leading to a non-uniform distribution of points of absorption on the absorbing sphere. The explicit forms of the $P_{1,2}(r)$ and $P_{1,2}(t|r)$ probabilities are given in Appendix B.

{\em Fully stochastic model}: An alternative way of accounting for the effects of the full mesoscopic evolution is to pre-compute the probability distributions of times for completion of the reaction, $P_r(t)$, and binding failure, $P_f(t)$. To compute these probabilities, an ensemble of trajectories that start at a uniformly chosen position on the inner sphere at radius $r_1$ is evolved until either the substrate binds and reacts or the unbound substrate escapes and passes through an absorbing sphere at intermediate distance $r_i$ from the binding site.  The binding probability can be estimated from the fraction of reactive trajectories and the probability densities $P_{r}(t)$ and $P_f(t)$ can be constructed using analytical fits to the estimated cumulative distribution functions obtained from the
reaction and failure times~\cite{VanZonSchofield08}. Given this information, once the substrate is at $r_1$ in step (4), the binding probability can be used to determine if reaction will occur and the reaction time can be drawn from $P_r(t)$ and used to complete the cycle, or if no reaction occurs the time can be drawn from $P_f(t)$ and used to increment the time.

\section{Simulation of PGK enzyme kinetics}\label{sec:sim}

The simulations employing hybrid MD-MPC dynamics were carried out on a system comprising a single PGK enzyme with bound ADP, a bPG substrate molecule and solvent molecules in a cubic box of length $L$ with periodic boundary conditions. The units used in the simulation are given in terms of length $\ell$, mass $m$, energy $\epsilon$ and time $\tau$. In these units the simulation box had length $L=40$ and contained $640,000$ solvent particles of mass $m=1$, resulting in a density $\rho = 10$.  The mass of the beads comprising the enzyme was taken to be $M=10$, so that the mass ratio of solvent to beads was set to $\mu = M/m = 10$. The solvent particles interact with all beads through the truncated repulsive potential in Eq.~(\ref{truncatedLJ}) with an adjustable $\sigma$, usually taken to be $\sigma = 1$. Simulation of the enzyme-substrate system consists of numerically integrating Newton's equations of motion for all bead and solvent particles that interact with a time step of $\Delta t=0.005$ for time intervals $\tau = 1$ between multiparticle collisions. Information from such direct simulations of the dynamics is required for both the diffusive encounters between the enzyme and substrate and the subsequent binding and reaction processes. These two aspects are discussed in the following subsections.

\subsection{Diffusive dynamics}
Although the diffusive encounters between the substrate and enzyme are treated analytically, these calculations require the diffusion coefficient $D$ of the substrate as input into the analytical formulas. Therefore, in this subsection we present results for $D$ for the explicit interaction and penetrating solvent models. Since the substrate does not interact with the enzyme in this regime we need only consider the motion of the substrate in pure solvent.

\begin{figure}[htbp]
\centerline{%
  \includegraphics[width=0.95\columnwidth]{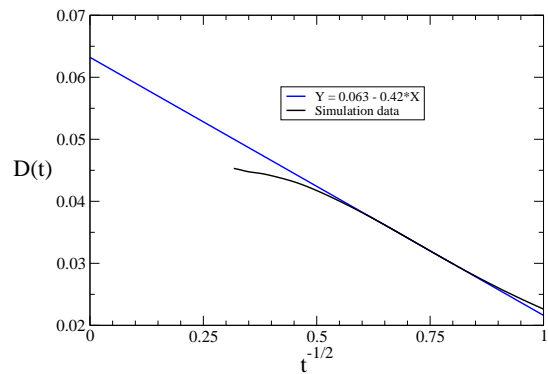}}
  \caption{The simulated value of the diffusion coefficient compared to the estimated time-dependent diffusion coefficient, $D(t)$, in Eq.(\ref{Dt}), versus $t^{-1/2}$ for an isolated Brownian particle with mass ratio = 10, $\rho = 10$, $k_B
  T= 1/3$, $\sigma = 0.5$.  From the fit of the data, the value of the diffusion coefficient is $D  = 0.063$.}
  \label{fig:diffusionSolvent}
\end{figure}

{\it Explicit interaction model}: In the explicit interaction model the substrate interacts with the solvent molecules through repulsive Lennard-Jones potentials and the solvent molecules undergo multiparticle collisions. The diffusion coefficient may be determined directly by simulation from the velocity autocorrelation function or the mean square displacement. Hydrodynamic effects are included in the MD-MPC dynamics and these give rise to long time tails in the velocity correlation function which make important contributions to the diffusion coefficient. For this reason it is convenient to estimate $D$ by extrapolation of the time-dependent diffusion coefficient to infinite time since
\begin{equation}
D(t) =\frac{1}{3}\int_0^t dt\; \langle {\bf V}(t)\cdot {\bf V} \rangle \sim D -\frac{\alpha_{_D}}{\sqrt{t}},
\label{Dt}
\end{equation}
where $\alpha_{_D} = (2/3) (4\pi (\eta+D))^{-3/2} (m\rho)^{1/2}$ with $\eta$ the shear viscosity. The power-law behavior of this quantity arises from coupling of the substrate to hydrodynamic modes of the solvent. The time-dependent diffusion coefficient is plotted versus $t^{-1/2}$ in Fig.~\ref{fig:diffusionSolvent} and shows the long-time power-law behavior. For a substrate with mass $M= 10$ in a solvent with $\rho = 10$, $k_BT= 1/3$, substrate-solvent Lennard-Jones parameters $\sigma = 0.5$ and $\epsilon =1$, we find $D  = 0.063$.
Hydrodynamic effects dominate the contributions to the diffusion coefficient and it is only weakly dependent on the mass of the substrate and solvent molecules. For a very large substrate molecule the diffusion coefficient takes a Stokes-Einstein form and is independent of the mass.

{\it Penetrating solvent model}: The diffusion coefficient can be computed analytically for the penetrating solvent model. In the collision step, the
rotation matrix is uniformly selected from a set of matrices in which the rotation by the angles $\alpha$ and $-\alpha$ around a given set
of axes are equally probable. The operation of the rotation
matrix on a general vector $\bm{r}$ for a rotation by angle $\alpha$ around a
unit vector $\hat{\bm{n}}$ can be written succinctly as
\begin{eqnarray}
\bm{\omega} \cdot \bm{r} = \bm{r} \cos\alpha + \hat{\bm{n}} (\hat{\bm{n}} \cdot
\bm{r})(1-\cos\alpha) + (\bm{r} \times \hat{\bm{n}}) \sin\alpha .
\end{eqnarray}
Since the substrate bead behaves as a point particle with respect to hydrodynamic flow, the only contribution to the self-diffusion coefficient comes from the rotation collision step.  Hence for this system, the decay of the velocity autocorrelation function for an isolated bead is expected to be a single exponential.

\begin{figure}[htbp]
\centerline{%
  \includegraphics[width=\columnwidth]{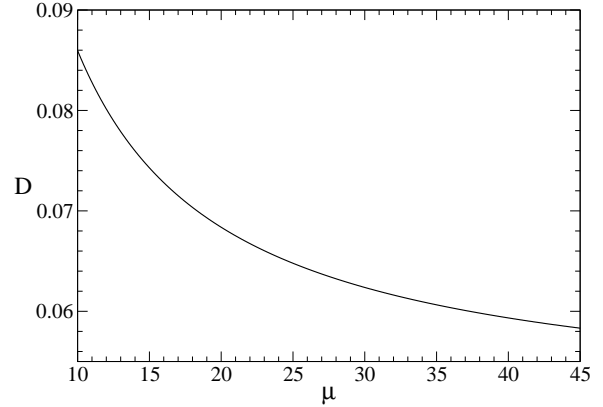}}
  \caption{Plot of the self-diffusion coefficient $D$ for a tagged
    particle in the penetrating solvent model as a function of the
    mass ratio $\mu$.}
  \label{fig:selfDiffusionPenetrating}
\end{figure}

The self-diffusion coefficient for this model can be computed from the velocity autocorrelation function using the trapezoidal rule,
\begin{eqnarray}
D &=& \frac{1}{3} \int_0^\infty dt \, \langle \bm{V} \cdot \bm{V}(t)
\rangle \\
&=& \frac{\tau}{3} \left( \frac{1}{2} \langle \bm{V} \cdot \bm{V}
  \rangle + \sum_{n=1}^{\infty} \langle \bm{V} \cdot \bm{V}(n\tau)
  \rangle \right),
\end{eqnarray}
where $\tau$ is the collision time and the brackets $\langle \cdots
\rangle$ correspond to an average over the stochastic realizations
(choice of rotation matrices) and the equilibrium distribution of the
system.  If the matrices are chosen uniformly and the rotation angles
$\alpha$ and $-\alpha$ are equally probable, then the Markovian
dynamics for a given cell has the limit distribution
\begin{equation}
P(n, \bm{r}_n, \bm{v}_n; R, V) = \frac{e^{-\rho}}{V_c^n}
\frac{\rho^n}{n!} \Pi_m(\bm{v}_n) \times \frac{1}{V_c} \Pi_m(\bm{V}),
\label{equilDist}
\end{equation}
where $n$ is the number of solvent particles in the cell containing the
tagged particle, $V_c$ is the volume of the cell (here taken to be unity) and $\Pi_m(\bm{v}_n)$ is the normalized
Maxwell-Boltzmann distribution for a system of $n$-particles at
temperature $T$.  Using this form, one finds that $\langle \bm{V}
\cdot \bm{V} \rangle = 3 k_BT/M$, and
\begin{eqnarray}
\langle \bm{V} \cdot \bm{V}(\tau) \rangle &=& \frac{1}{n_R}
\sum_{i=1}^{n_R} \langle \bm{V} \cdot \left( \bm{v}_c + \bm{\omega}_i
  \cdot (\bm{V} - \bm{v}_c ) \right) \rangle \nonumber\\
&=& \langle \bm{V} \cdot \bm{v}_c \rangle + \langle \bm{V} \cdot
\overline{\bm{\omega}} \cdot \left( \bm{V} - \bm{v}_c\right) \rangle ,
\end{eqnarray}
where $\overline{\bm{\omega}} = \sum_{i=1}^{n_R} \bm{\omega}_i /n_R$
and $n_R$ is the total number of rotation matrices.  Inserting the
stationary density in Eq.~(\ref{equilDist}), and defining the mass
ratio $\mu = M/m$, one gets
\begin{eqnarray}
&&\langle \bm{V} \cdot \bm{V}(\tau) \rangle =
\frac{3k_BT}{M} e^{-\rho} \sum_{n=1}^{\infty} \frac{\rho^n}{n!}
\left( 1 + (c_\gamma -1) \frac{n}{n+\mu} \right) \nonumber \\
&&\qquad \qquad= \frac{3k_BT}{M} \left( 1 + \frac{(c_\gamma - 1) \rho }{1+\mu}
  M(1,2+\mu,-\rho) \right) \nonumber \\
&& \qquad  \qquad \equiv \frac{3k_BT}{M} (1 - \gamma),
\end{eqnarray}
where $M(1,2+\mu,-\rho)$ is Kummer's
function of the first kind\cite{AbramowitzStegun} and $c_\gamma =
\text{Tr} \, \overline{\bm{\omega}}/3$.  If there is no correlation between
solvent particles occupying the cell containing tagged particles following the collision steps, so that $\langle \bm{V} \cdot \bm{V}(n\tau) \rangle =
(1-\gamma) \langle \bm{V} \cdot \bm{V}((n-1)\tau) \rangle$, we
conclude
\begin{equation}
D = \frac{k_BT \, \tau}{M} \left( \frac{1}{2} + \sum_{n=1}^{\infty}
  (1-\gamma)^n \right) =\frac{k_B T \, \tau}{M} \left( \frac{2-\gamma}{2\gamma} \right),
\label{Dpenetrating}
\end{equation}
where
\begin{equation}
\gamma = \frac{1-c_\gamma}{1+\mu} \, \rho \, M(1,2+\mu, - \rho).
\end{equation}
The self-diffusion coefficient is plotted in Fig.~\ref{fig:selfDiffusionPenetrating} as a function of the mass ratio $\mu$ for the simulation values $k_BT = 1/3$ and $\rho = 10$ and $c_\gamma = 1/3$.

While the diffusion coefficient depends weakly on the mass ratio for the explicit solvent interaction model, it does depends strongly on the mass ratio for the penetrating solvent model. In order to facilitate comparisons between these two solvent interaction models, we choose the mass ratio so that the self-diffusion coefficient of an isolated bead matches that in the interacting solvent model. Note that for the mass ratio $\mu = 10$ used in the interacting solvent model, the self-diffusion coefficient in the penetrating solvent model is substantially larger than in the interacting model ($D = 0.086 > 0.063$), and a mass ratio of roughly $\mu = 28.5$ must be used for the dynamics of the tagged particle to be comparable.  Simulations of
an isolated Brownian particle immersed in the penetrating solvent validate the predictions of Eq.~(\ref{Dpenetrating}). Finally, we note that the penetrating solvent model without hydrodynamic interactions is also given by Eq.~(\ref{Dpenetrating}).

\subsection{Substrate binding and reaction}
The position of the bPG substrate was randomly chosen on a spherical shell at a distance $r_{1}=7$ from the active binding site of the enzyme.  The distance was chosen so that the bPG substrate does not interact with the active site or other parts of the enzyme.  For each realization of the dynamics, the enzyme configuration was equilibrated in the presence of the solvent while constraining the bPG substrate in position.  The run was then initiated by randomly drawing the bPG velocity from a Maxwell-Boltzmann distribution at an effective temperature of $k_BT = 1/3$  and releasing the constraint.  If the substrate bound to the enzyme (determined by a distance criterion), the time of binding of the bPG was recorded.  If instead the distance of the substrate to the active site reached a large value, here taken to be at a substrate-active site distance of $r_i = 9$, the evolution of a realization was terminated and the failure time was recorded.  Upon binding, the form of the network potential for the enzyme allows the enzyme to close to an activated form.  The time of closing, again determined by a distance criterion between conserved, rigid sections of the enzyme, was recorded.  Once the enzyme closed, a reaction time $\tau_r$ was drawn from a Poisson distribution (here taken to have a mean reaction time of $\overline{\tau}_r=25$ time units), which defines the rate at which an unbinding potential was activated by the control parameter $\xi$.

\begin{figure}[htbp]
\centerline{%
  \includegraphics[width=\columnwidth]{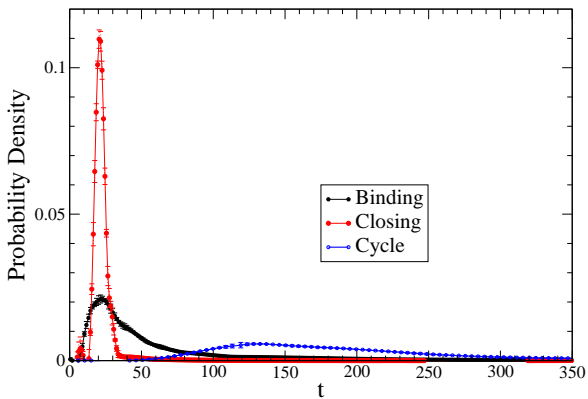}}
  \caption{Probability densities for the full solvent model as a
    function of the collision time.  The results are for simulation
    conditions $\mu = 10$, $k_BT = 1/3$, $\rho = 10$, with a solvent-bead
    interaction $\sigma = 0.5$ cell lengths, corresponding to
    $\sigma = 2.5 \AA$.}
  \label{fig:fullHydroDensity}
\end{figure}

The probability densities for the time of substrate binding, the closing time of the enzyme after binding, and the overall cycle time are shown in
Fig.~\ref{fig:fullHydroDensity}.  The analytical fit to the densities with bootstrap estimates for uncertainties were computed from the raw data using the procedure described in Ref.~[\onlinecite{VanZonSchofield10}].
A prominent feature in the probability density of binding times is the
long algebraic tail, which is a signature of the substrate initially
moving away from the enzyme but eventually diffusing into the active
site.  The form of the tail in this density is consistent with the
asymptotic long time behavior for a particle diffusing into an
absorbing region in three dimensions.  Note that the probability
density for the overall cycle time can be decomposed into a
convolution of the density for binding, closing and diffusion away from
the binding site after the reaction is complete.  Since diffusive
motion leads to densities with heavy tails, the overall cycle time
density is broad, which is characteristic of algebraic tails.

Another important qualitative feature of the solvent-enzyme model is
the variable degree of solvation of the bPG substrate during the
binding process.  When the distance $\sigma = 0.5$ characterizing the
solvent-bead repulsion is large enough, the solvent is unable to
penetrate the volume occupied by the enzyme.  The bPG substrate binds
to a region inside the enzyme that is exposed when the enzyme is in an
open conformation.  Upon binding, the enzyme closes via a hinge-like
mechanism and brings the ADP-bPG substrates near one another enabling
the transfer of the phosphoryl group.  Less solvent is able to
penetrate into the binding pocket of the bPG substrate in the closed conformation of
the enzyme, and hence solvent is expelled from the pocket as bPG binds
and the enzyme closes, providing a favorable environment for the catalysis.~\cite{banks79,inoue10}

\begin{figure}[htbp]
\centerline{\includegraphics[width=\columnwidth]{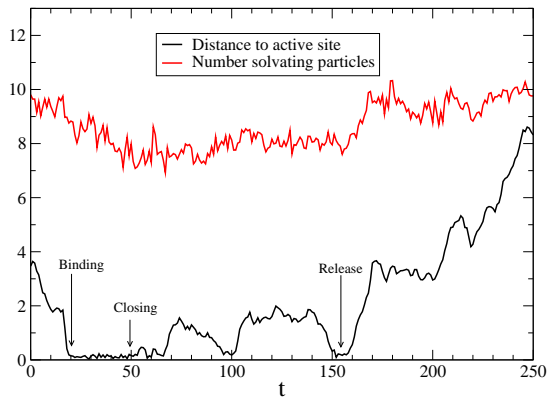}}
\centerline{\includegraphics[width=\columnwidth]{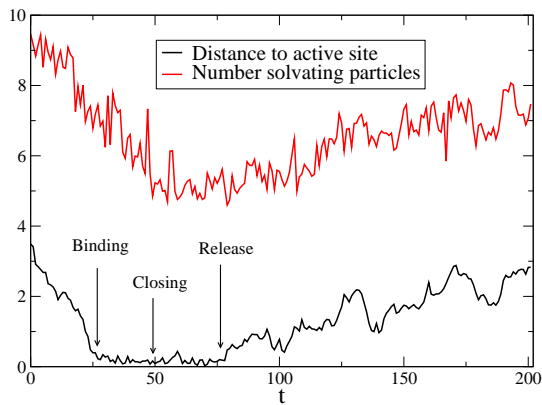}}
  \caption{Time series showing the reduction in the number of solvent particles in the vicinity of the bPG
    substrate as it binds to the enzyme.  The red curves show the number of solvent particles in the
    cell containing the bPG substrate as a function of time, while the black curves denote the distance of the substrate to the enzyme binding site (measured in cell length units, where 1 cell length is $5 \AA$). (top) $\sigma = 0.5$, (bottom) $\sigma = 0.7$.}
  \label{fig:drying}
\end{figure}

The expulsion of solvent can be tracked by
computing the local solvent density around the bPG substrate as it binds
and reacts, as can be seen in Fig.~\ref{fig:drying}.  This drying effect is highly sensitive to the
choice of the repulsive interaction parameter $\sigma$.
When $\sigma = 0.5$ (see top panel of Fig.~\ref{fig:drying}), the bound substrate typically has $2$ fewer
solvent particles solvating it, whereas away from the enzyme the
average number of solvating fluid particles corresponds to the value
of the bulk density ($\rho = 10$). This difference between bulk and bound
solvation levels increases as the repulsion parameter $\sigma$ increases (see bottom panel of
Fig.~\ref{fig:drying} where $\sigma = 0.7$).  There are important differences in the
qualitative nature of the dynamics when the repulsion parameter becomes
large.  Although the exterior of the enzyme experiences a larger
overall friction, the dissipating effect of the solvent on the
enzyme-substrate interaction is decreased in the pocket of the enzyme
where the binding occurs.  The bPG substrate retains a high kinetic
energy upon entering the pocket for a longer period of time due to a
limitation in the simple model of the binding process in which the substrate effectively interacts
with only a few beads of the enzyme.  Because of the limited coupling
of the beads in the active site to other beads in the protein,
the excess energy of the substrate is slowly dispersed into internal
motions of the protein and solvent.  For this reason, we focus
primarily on a regime in which the solvent rapidly dissipates energy
($\sigma = 0.5$).

\begin{figure}[htbp]
\centerline{%
  \includegraphics[width=\columnwidth]{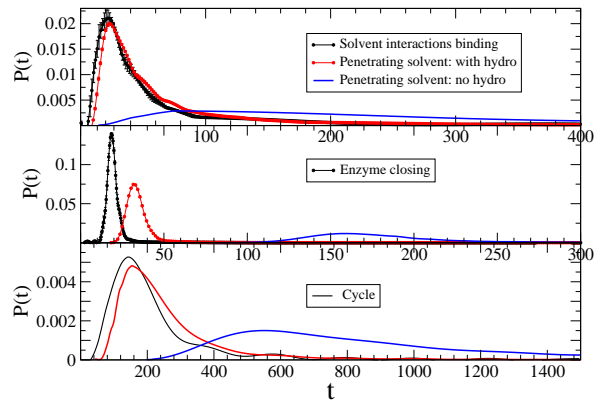}}
  \caption{Probability densities $P(t)$ for substrate binding
    (top panel), enzyme closing (middle panel), and total reaction
    cycle (bottom panel) versus time.
    The black curves correspond to results for the interacting solvent
  model, the red curves correspond to the results for the penetrating
  solvent model with hydrodynamics and the blue curves are the results
for the penetrating solvent model without hydrodynamics.}
  \label{fig:distributions}
\end{figure}
In Fig.~\ref{fig:distributions} the probability densities for
the binding time, enzyme closing time and overall cycle time are
presented.  Looking at the top panel, we see that the probability
densities of the binding time for the interacting and penetrating solvent
models are comparable once the dynamics has been properly scaled by
the mass ratio.  This similarity is not surprising, as the time scale
for binding is primarily determined by diffusive motion and is not
sensitive to the level of solvation of the substrate by the fluid
particles.  However the absence of hydrodynamic flow around the enzyme
and substrate has a profound effect on both the form of the probability
density, which is significantly broadened, and the mean binding time,
which is shifted by a factor of roughly  a factor of three. In addition, the binding probability is significantly reduced from $P_r=0.078$,
in the presence of hydrodynamics, to $P_r=0.03$, which can have a significant impact
on the density for the overall substrate conversion time when the
concentration of substrates is elevated. Note that  the probability density of
binding times has a strong tail for all models, indicative of the
importance of the diffusive dynamics experienced by the substrate.

The time required for the enzyme to close after binding is noticeably
different in all three models.  The penetrating solvent model does not
account for solvent expulsion as the enzyme closes, and therefore has
a higher net friction and longer time scale than is present in the
explicit interaction model.  Once again, the effect of hydrodynamics is
significant, and shortens the time required for the enzyme
to close.

The overall cycle time density is a convolution of the binding time and
closing time densities, and is therefore different for all three models.

\subsection{Fully stochastic model}
A stochastic procedure can be implemented for the overall enzymatic
process using data from the numerical simulations and the computed values of the binding
probability starting from a radial distance of $r_1$.  If the binding
is accepted starting from the inner sphere with probability
$P_r$, which for the explicit solvent model is approximately $P_r= 0.078$,
the overall cycle time for the reactive process can
be added to the overall time for the process by drawing from the
numerically-obtained probability densities and cumulative distributions.
To carry out
the procedure, the reaction time is drawn by numerically solving the equation
$C_{\text{cycle}}(t_u) = u$ for the time $t_u$ using bisection or
Newton-Raphson methods, where $u$ is a
random variable drawn uniformly from the unit interval.  Here,
$C_{\text{cycle}}(t)$ is the cumulative distribution for the cycle
obtained from the simulation.

To convert the system collision time into physical units, note that
the self-diffusion coefficient in system units is $0.06$ $\ell^2/\tau$.
Equating this with the desired value of the diffusion coefficient in
the cytoplasm of roughly $D= 1 \cdot 10^{-6} \, \text{cm}^2/\text{s}$,
we conclude that $\tau = 1.5 \cdot 10^{-10}$ seconds.
Using this scaling, we find that the typical time required for the PGK
enzyme to close following binding of both substrates is on the order of
$3$ to $6$ $ns$ for the solvent models incorporating hydrodynamic flow,
which is consistent with experimental~\cite{inoue10,haran92} and simulation~\cite{palmai09} studies of the enzyme domain motions.

\begin{figure}[htbp]
  \includegraphics[width=\columnwidth]{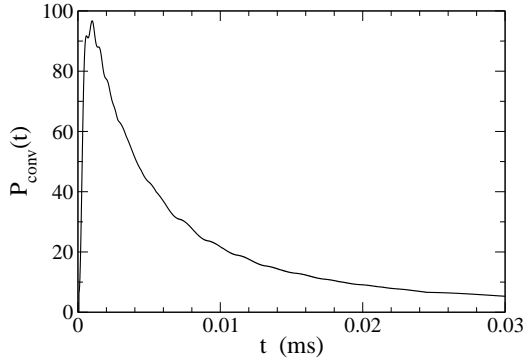}
  \caption{Probability density $P_{\rm{conv}}(t)$ of the substrate conversion time to products versus time expressed in milliseconds
    for the explicit solvent model.  The other models yield essentially identical results since the
    substrate conversion is determined primarily by diffusion when the substrate is
    at physiological concentrations}
  \label{fig:turnover}
\end{figure}

The probability density $P_{\rm{conv}}(t)$ of substrate conversion times is
shown in Fig.~\ref{fig:turnover}.
Somewhat surprisingly, no difference in the probability density of substrate conversion
time is readily observable at the enzyme concentration studies here even though the
binding probability is more than two times larger in the presence of hydrodynamics than
in its absence.  This is due to the multiple convolutions of the first passage time densities
which have heavy and prominent tails that tend to smooth out observable
differences after multiple convolutions.

\section{Summary} \label{sec:conc}
A stochastic method for computing the probability density of the time required for the enzymatic catalysis of a substrate to product was constructed. The method consists of combining analytical computations of binding probabilities and first-passage times of a substrate diffusing between two concentric
absorbing spheres with explicit simulation of motion of the substrate in the immediate vicinity of the enzyme.  Once the explicit simulations have been performed and the data analyzed in terms of binding probabilities and first passage time densities, the method allows the probability density of the time required for the phosphate transfer to be computed at a variety of enzyme concentrations.

The method was illustrated by considering the catalytic transfer of a phosphate group from bPG to a bound ADP substrate by the phosphoglycerate kinase enzyme under physiological conditions.  The binding probability and phosphoryl group transfer times for a substrate diffusing in a $0.1$ mM concentration of phosphoglycerate kinase were computed under three different solvent conditions using a network model of the enzymatic system constructed from the morphing analysis of the conformational change between the open and closed conformations\cite{pgk1,pgk2} of the enzyme.  The solvent models were chosen to selectively account for various degrees of correlated solvent motion to probe the importance of collective flow effects on the enzyme dynamics.  It was
demonstrated that dynamical solvent flow effects assist the binding of the substrate to the active site
of the enzyme and facilitate the hinge motion of the enzyme that leads to its closing. Two different models that incorporate hydrodynamic flow effects,
one with direct solute-solvent interactions and another penetrating solvent model where solvent particles are treated as point particles in their interactions with the substrate and protein, have similar binding probabilities and cycle time densities. However, the density profiles of the solvent near the active site
as the enzyme closes post-substrate binding differ, since expulsion of the solvent from the binding pocket is not possible for the penetrating-solvent model.
In contrast, a Smoluchowski-type model in which all beads feel a friction that is independent of the conformation of the enzyme is characterized by a lower substrate binding probability and a shift in the cycle time density to larger time scales relative to the models incorporating hydrodynamic
effects.  The lower substrate binding probability leads to a detectable shift in the maximum appearing in the density of substrate conversion times.

The validity of the stochastic method presented here relies on a number of assumptions that are
questionable for the behavior of the enzymatic system in a cellular environment.  It has been assumed
that the enzymes are homogeneously distributed with no correlation between their positions in the volume.
It is quite possible that the enzymes are, in fact, locally clustered in the cytoplasm in a way that effectively
reduces the distance between them and the substrates thereby enhancing their efficiency.  This is likely to
be the case if there is correlation between the spatial location of the phosphoglycerate kinase enzyme
and enzymes such as glyceraldehyde phosphate dehydrogenase that act earlier in glycolysis.  In addition, it has been
assumed that the dynamics of the substrate in the complex, crowded cytoplasm is diffusive, which may
be reasonable on long time scales but less accurate on the time scale of solvent motion. However, subdiffusive motion of
proteins and finite-size probe molecules has been seen in crowded cellular environments.~\cite{wachsmuth00,weiss04,banks05,guigas07} Nonetheless,
assuming substrates do move diffusively in the cytoplasm at long times,
the diffusive nature of the substrate dynamics leads to a broad distribution of substrate conversion times
that differs substantially for the exponential distribution one might anticipate from mass action kinetics.

It is straightforward, though computationally intensive, to incorporate more detailed models of the
enzymatic system to produce quantitatively accurate results.  This is readily accomplished by performing
all atom simulations of the system complete with detailed molecular mechanical-based interaction
potentials and quantum-mechanical analysis of chemical reaction pathways.  Nonetheless,
it is likely that the observation that the solvent flow assists the binding and subsequent protein
motions will also be observed in more detailed models of the enzymatic system.

\begin{acknowledgments}
Computations were performed on the GPC supercomputer at the SciNet HPC Consortium, which is funded by the Canada Foundation for Innovation under the auspices of Compute Canada, the Government of Ontario, the Ontario Research Fund Research Excellence and the University of Toronto.

This work was supported in part by grants from the Natural Sciences and Engineering Council of Canada.  The authors would like to Dr. Ramses van Zon for useful discussions.
\end{acknowledgments}

\section*{Appendix A: PGK potential functions}

In this Appendix we give the detailed form of the potential function $V_{PS}$ that governs the dynamics of the protein and its interactions with the bPG substrate.

Bonds in the set ${\mathcal B}_c$ of common links were assigned bond potentials $V_c(r_{ij})$ constructed in the following way. The potentials for the common links in the open and closed configurations of the enzyme, $V_{co}$ and $V_{cc}$, are given by
\begin{eqnarray}\label{commonPotential}
&&V_{co, cc}  = \frac{k_h}{2} \sum_{<ij> \in {\mathcal B}_{hc}} \left( R_{ij} - l_{ij}^{(o,c)} \right)^2 \\
&&\qquad + \epsilon \sum_{<ij> \in {\mathcal B}_{sc} } \left( 5 \left( \frac{\sigma_{ij}^{(o,c)}}{R_{ij}} \right)^{12} - 6
\left( \frac{\sigma_{ij}^{(o,c)}}{R_{ij}} \right)^{10} \right) ,\nonumber
\end{eqnarray}
where the parameters $l_{ij}^{(o,c)}$ and $\sigma_{ij}^{(o,c)}$ were determined by the equilibrium distances for the harmonic and soft-common links in the open and closed conformations and $k_h$ is the force constant for the hard elastic network bonds. Given this input, the potential for the common interactions $V_c$ was taken to be the lowest eigenvalue of a two-dimensional empirical valence bond (EVB) matrix with constant off-diagonal elements $\Delta$, so that~\cite{maragakis05}
\begin{equation}
V_c = \frac{1}{2} \left(  (V_{co} + V_{cc}) - \left( (V_{co}-V_{cc})^2 +
4\Delta^2 \right)^{1/2} \right).
\end{equation}
This form of the potential allows the system to smoothly switch between stable open and closed configurations. Links in the soft-open, ${\mathcal B}_{so}$, and soft-closed, ${\mathcal B}_{sx}$ sets were assigned bond potentials
\begin{equation}
V_s(R_{ij})=\epsilon \Big( 5 \Big( \frac{\sigma}{R_{ij}}
\Big)^{12} - 6 \Big( \frac{\sigma}{R_{ij}} \Big)^{10} \Big)
\end{equation}
with identical forms.
In addition, monomeric beads representing amino acid residues repel one another at short distances according to a truncated Lennard-Jones (LJ) potential
\begin{equation}\label{truncatedLJ}
V_r = \sum_{<ij>}\epsilon_{bb} \left( \left( \frac{\sigma_{bb}}{R_{ij}} \right)^{12} -
2 \left( \frac{\sigma_{bb}}{R_{ij}} \right)^{6} +1\right) \theta(\sigma_{bb}-R_{ij}),
\end{equation}
where $\theta(x)$ is the Heaviside function.  The bPG substrate, represented by a single bead with coordinate ${\bf R}$, also interacts with all beads in the protein through a repulsive LJ potential of this form, $V_{r}^{{\rm (b)}}(R_{bi})$, where $R_{bi}=|{\bf R}- {\bf R}_i|$ with $\epsilon_{bs}$ and $\sigma_{bs}$ energy and distance parameters.

\subsection*{Interactions governing the reactive event and conformational changes}
The binding interaction $V^{{\rm (b)}}_{b}({\bf R},{\bf R}_0^a, {\bf R}_1^a, {\bf R}_2^a)$ between the bPG substrate at position ${\bf R}$ and the enzyme was designed to depend on the distance between the substrate and bead with coordinate ${\bf R}_1^a$, as well as the orientation of the substrate with respect to a coordinate frame determined by three beads defining the binding pocket of the enzyme. Defining the relative position vector $\bm{R}_{S1} = \bm{R} - \bm{R}_1^a = R_{S1} \hat{\bm{R}}$ with magnitude $R_{S1}$ and direction $\hat{\bm{R}}$ of the substrate with respect to a coordinate system centered on the binding site $\bm{R}_1^a$, the projection $R_{S1}^z = \hat{\bm{R}} \cdot ( \hat{\bm{R}}_{10}^a \times \hat{\bm{R}}_{21}^a)$ is computed,
where $\hat{\bm{R}}_{ij}^a$ is the unit vector along $\bm{R}_{ij}^a = \bm{R}_i^a - \bm{R}_j^a$.  The binding potential is then taken to be
\begin{eqnarray}\label{bpgBind}
&&V^{{\rm (b)}}_{S} = f(R_{S1}) \Bigg[
\epsilon \Bigg[ \left( \frac{\sigma_{bb}}{R_{S1}} \right)^{12} -
\left( \frac{\sigma_{bb}}{R_{S1}} \right)^{6} -
3 \left( \frac{\sigma_{bb}}{R_{S1}} \right)^{2} \Bigg] \nonumber\\
&& \qquad +
K_{S} \left( \frac{\sigma_{bb}}{R_{S1}} \right)^{12}
\left( 1-(R_{S1}^z)^2 \right) \Bigg] \theta(-R_{S1}^z)
\end{eqnarray}
where $K_{S} = 1.5$ in the energy units. In Eq.~(\ref{bpgBind}), $f(R)$ is a smooth cut-off function
\begin{equation}
f(R) = \left\{
\begin{array}{ll}
1, & \mbox {$R < R_\ell$}\\
\frac{(R_u - R)^2}{(R_u-R_\ell)^3} \left( R_u - 3 R_\ell + 2R \right),
& \mbox{$R_\ell \leq R \leq R_u$} \\
0, & \mbox{$R > R_u$}
\end{array}
\right. ,
\end{equation}
where the upper and lower cut-off values are set to $R_u = 3 \sigma$ and
$R_\ell = 2.5 \sigma$.  The potential insures that the optimal angle of approach and binding of the substrate in the active site pocket is along the $\hat{\bm{R}}_{21}^a \times \hat{\bm{R}}_{10}^a$ direction.
In principle, the excluded volume interactions of the
substrate bead with the enzyme beads are sufficient to determine the binding
pathway of the substrate, while the orientational dependence of the binding
potential in Eq.~(\ref{bpgBind}) restricts the binding location in the active site.

As the substrate binds it triggers conformational changes in the protein that lead to hinge closing to bring the bPG and ADP substrates into proximity for the phosphoryl group transfer.   Thus, as bPG interacts with the protein in the course of binding to the active site, the open protein configuration is destabilized with respect to the closed configurations, driving the enzyme towards the closed conformation. To achieve this conformational change in the network model, the interaction potentials for the soft, non-common set of links are modified. We define the reaction coordinate $\xi$, where
\begin{equation}
\xi = \frac{1}{2} \left( 1 + \tanh{x} \right),
\end{equation}
where
\begin{equation}
x = \frac{ (R_{b1} - R_{b1}^{o})^2}{(R_{b1}-R_{b1}^{c})^2} -
\frac{ (R_{b1} - R_{b1}^{c})^2}{(R_{b1}-R_{b1}^{o})^2} ,
\end{equation}
and $R_{b1}^o$ is the initially large distance between the substrate and the binding site in the enzyme in the open configuration and $R_{b1}^c$ is the same distance in the bound, closed complex.  Since the substrate is unbound and hence far from the binding pocket in the starting configuration, $R_{b1}^o \gg R_{b1}^c$.   Note that when the substrate is far from the enzyme, $x$ is large and negative and $\xi \approx 0$, whereas $x$ becomes large and positive as the substrate moves towards the binding site with the result that $\xi \approx 1$ upon binding. Given this reaction coordinate, the soft, non-common potential function is taken to be
\begin{eqnarray}
V_{nc} = \xi \sum_{<ij> \in {\mathcal B}_{sx}} V_s(r_{ij}) +  (1-\xi) \sum_{<ij> \in {\mathcal B}_{so}} V_s(r_{ij}).
\end{eqnarray}

The protein-substrate interaction potential is given by the sum of these contributions:
\begin{equation}
V_{PS}({\bf R}^{N_P},{\bf R}; \xi(R_{S1}))=  V_c+V_r+V_{r}^{{\rm (b)}}+V^{{\rm (b)}}_{S}+V_{nc}.
\end{equation}

After binding, the reaction coordinate $\xi$ is treated as an external control parameter that is governed by the equation:
\begin{equation}
\xi (t) = \left\{
\begin{array}{ll}
1 - t/\tau_r & \mbox{if $t \leq \tau_r$} \\
0 & \mbox{otherwise},
\end{array}
\right.
\end{equation}
where $\tau$ is the reaction time drawn from an exponential distribution $P(\tau_r ) = \overline{\tau}_r^{-1} e^{-\tau_r/\overline{\tau}_r}$ and $\overline{\tau}_r$ is the average reaction time.  Upon completion of the reaction when $\xi = 0$, the interaction between the substrate in the binding pocket and the binding site is changed to a repulsive Lennard-Jones interaction to reflect the unstable interaction of the altered substrate and the binding pocket.  Since $\xi = 0$, the closed configuration is unstable and the enzyme reopens, completing the cycle.  In this treatment, the reaction is treated irreversibly and the surrounding solvent absorbs energy from the chemical process, leading to a slight heating of the solvent.  The average reaction time $\overline{\tau}_r$ is taken to be $25$ time units, corresponding to a physical reaction time of roughly $2.5$ $ns$.  Note that the precise value for the average reaction time is unimportant for looking at the qualitative effects
of the solvent environment on the dynamics of the enzymatic system. Detailed quantum chemical calculations are required to determine if this estimate of the reaction time from the metastable bound state to a final state consisting of the products bound in a closed conformation of the enzyme is reasonable.

\section*{Appendix B: Diffusion of substrate to a region near enzyme}
The first passage time distribution $P(t|\bm{r}_0)$ for a Brownian walker
starting from position $\bm{r}_0$ at time $t=0$ onto a sphere centered
at the origin can be computed
from the survival probability distribution $F(t|\bm{r}_0)$ using
\begin{equation}
P(t|\bm{r}_0) = - \frac{dF(t|\bm{r}_0)}{dt}\bigg|_{\mathrm sphere} ,
\end{equation}
where the derivative of $F(t)$ only includes the flux of walkers into
the sphere and
\begin{equation}
F(t|\bm{r}_0) = \int_{\Omega} d\bm{r} \, P(\bm{r}, t; \bm{r}_0).
\label{survival}
\end{equation}
In Eq.~(\ref{survival}), $P(\bm{r}, t; \bm{r}_0)$ is the conditional
probability of finding the walker at position $\bm{r}$ at time $t$
given that it was initially at $\bm{r}_0$, and $\Omega$ is the domain
of the system.  We shall assume that the walker is confined between
two absorbing spheres of radii $r_{-}=r_1$ and $r_{+}=r_2$.  Given the
spherical boundaries of the domain, it is natural to express positions
in terms of spherical polar coordinates $(r, \theta, \phi)$, where the
$z$-axis from which the angle $\theta$ is measured relative to the
vector connecting the origin to a specific point $\bm{r}_p$ on the
inner sphere.  The angle $\phi$ can measured from the plane containing
the vectors $\bm{r}_p$ and $\bm{r}_0$ so that $\phi_0 = 0$.  The evolution of
the conditional probability is determined by the diffusion equation
\begin{equation}
\frac{\partial P}{\partial t} = D \nabla_r^2 P,
\end{equation}
and satisfies the boundary condition $P(\bm{r},0;\bm{r}_0) = \delta
(\bm{r} - \bm{r}_0)$.  From the diffusion equation,
we find that the first-passage distribution through a spherical domain
at radial distance $r_{-}$ is given by
\begin{equation}
\begin{split}
P(t|\bm{r}_0) &= - \int_{\Omega} d\bm{r} \, \frac{\partial P}{\partial t} = -D
\int_{\Omega} d\bm{r} \, \nabla_r^2 P \\
&= -D \int_{\partial \Omega} dS \; \hat{\bm{r}} \cdot \nabla_{\bm{r}} P \\
&=
D \int_{\mbox{sphere at $r_{-}$}} dS \, \frac{\partial P}{\partial r} ,
\end{split}
\end{equation}
where the second line follows from Green's theorem.  The domain
$\Omega$ contains all points with radii in the range $[r_{-},r_{+}]$,
and the integral over the inner sphere can be written to obtain
\begin{equation}
P(t|\bm{r}_0) = D r_{-}^2 \int_{0}^{\pi} d\theta \sin\theta \, \int_{0}^{2\pi}
d\phi \, \frac{\partial P(r_{-},\theta, \phi, t; \bm{r}_0)}{\partial r}.
\end{equation}

The diffusion equation may be solved for arbitrary coordinates
$\bm{r}$ and $\bm{r}_0$ in the presence of absorbing boundaries
by expanding the
density $P(\bm{r},t|\bm{r}_0)$ in spherical polar coordinates.  The absorbing boundary
conditions require that
\begin{equation}
\begin{split}
P(r_{-}, \theta, \phi, t; r_0, \theta_0, \phi_0) &= 0 \\
P(r_{+}, \theta, \phi, t; r_0, \theta_0, \phi_0) &= 0.
\label{absorbBoundary}
\end{split}
\end{equation}
Although a general series solution in spherical harmonic functions for $P(\bm{r}, t |\bm{r}_0)$ is possible,
the spherically averaged flux $F(t|\bm{r}_0)$
and first-passage time distribution $P(t|\bm{r}_0)$ are simple to obtain since only
the first, spherically-symmetric term in the expansion remains. From the
differential equation for the expansion coefficients,  one finds that the
Laplace transform
$\tilde{P}(s|\bm{r}_0) = \int_0 ^{\infty} dt \,  e^{-{st}} P(t|\bm{r}_0)$ of the first passage time density $P(t|\bm{r}_0)$ for the inner sphere is given by\cite{Redner}
\begin{eqnarray}
\tilde{P}(s|\bm{r}_0) &=& \left( \frac{x_{-}}{x_0} \right)^{1/2}
\frac{C_{1/2}(x_0,x_{+})}{C_{1/2}(x_{-}, x_{+})} \nonumber \\
&=&  \left( \frac{r_{-}}{r_0} \right)^{1/2}
\frac{C_{1/2}(x_0,x_{+})}{C_{1/2}(x_{-}, x_{+})} ,
\label{LaplacePassage}
\end{eqnarray}
where  $x_0$ is the scaled variable $x_0=\sqrt{s/D} \, r_0$,
$x_+ = \sqrt{s/D} r_+$, $x_{-} = \sqrt{s/D} r_{-}$, and
$C_\nu (a,b) = I_{\nu}(a)K_{\nu}(b) - I_{\nu}(b) K_{\nu}(a)$, where $I_\nu (x)$ and $K_\nu (x)$ are modified Bessel functions.
For a large outer sphere for which $r_{+} \gg r_{-}$,
$C_{1/2}(x,x_{+}) \rightarrow -I_{1/2}(x_{+}) K_{1/2}(x)$.
Considering a particle that can start at any point on a spherical shell
at $r=r_0$, we can write $\tilde{P}(s|\bm{r}_0) \sim \sqrt{r_{-}/r_{0}} \; K_{1/2}(x_0)/K_{1/2}(x_{-})$.

\begin{figure}
\centerline{\includegraphics[width=\columnwidth]{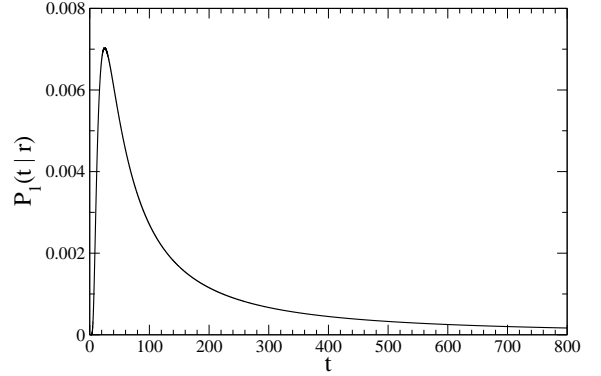}}
\centerline{\includegraphics[width=\columnwidth]{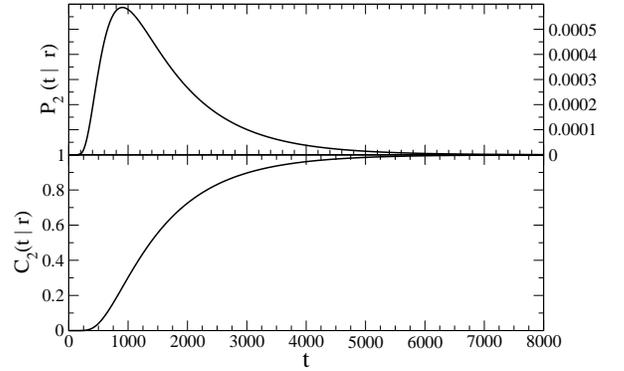}}
  \caption{Absorption time probability density versus time.  The top
  panel is the absorption time for the absorption onto an inner sphere
  at $r_1=7$ starting from a radial distance $r = 10$ in length units $\ell$.
  The bottom panel shows the absorption time density (top) and cumulative distribution (bottom) for the outer sphere, where the
  outer absorbing sphere radius is set to be $r_2=31.6$ and $r=10$.}
  \label{fig:absorption}
\end{figure}

Noting that
\begin{eqnarray*}
k_0(x) = \sqrt{ \frac{\pi}{2x}} \, K_{1/2}(x) = \frac{\pi}{2x} e^{-x},
\end{eqnarray*}
the Laplace transform $\tilde{P}_1(s|r_0)$ of the first passage density to the inner
sphere can be approximated by
\begin{equation}
\tilde{P}_1(s|r_0) = \frac{k_0(x_0)}{k_0(x_{-})} = \left(\frac{r_{-}}{r_0} \right)
e^{-\sqrt{s/D} \, \big( r_0-r_{-} \big)},
\end{equation}
which can be explicitly inverted to obtain the normalized first-passage
distribution $P_1(t|r)$ for particles that are absorbed at the inner sphere radial distance $r_1$ starting from the
spherical shell at distance $r$,
\begin{equation}
P_1(t|r) = \frac{ \big( r -r_{1} \big)}{\sqrt{4\pi D t^3}} e^{-
\left( r -r_{1} \right)^2/(4Dt)}.
\label{firstPassageSphere}
\end{equation}
This result is plotted in Fig.~\ref{fig:absorption} (top panel).
Note that the fraction of particles absorbed at the inner sphere in the infinite time limit can be computed from the $s=0$ limit of Eq.~(\ref{LaplacePassage}), yielding
\begin{eqnarray*}
\tilde{P}_1(s=0|r) = P_1(r) = \frac{r_{1}}{r} \frac{r_2-r}{r_2 - r_1}.
\end{eqnarray*}
The fraction of particles absorbing at the outer boundary in the infinite
time limit is
$P_2(r) = 1- P_1 (r)$.  These probabilities play an
important role in the stochastic simulation algorithm.

The first-passage time density at the outer sphere is obtained
similarly, although the inversion of the Laplace transform $\tilde{P}_2(s|r)$ is complicated since
\begin{eqnarray*}
\tilde{P}_2(s|r) &=&  \left( \frac{r_{2}}{r} \right)^{1/2}
\frac{C_{1/2}(x ,x_{2})}{C_{1/2}(x_{1}, x_{2})} \\
&=& \frac{r_2}{r} \frac{\sinh x_1 \, e^{-x} - \sinh x \, e^{-x_1}}{\sinh x_1
  \, e^{-x_2} - \sinh x_2 \, e^{-x_1}},
\end{eqnarray*}
where $x = \sqrt{s/D} r$ and $x_i = \sqrt{s/D} r_i$.  Although the
density can be approximated using series expansions for $\Theta$-functions, it is a simple matter to invert $\tilde{P}_2(s|r)$
numerically using the Stehfest algorithm\cite{Stehfest70a,Stehfest70b}.

To draw a random time from the first-passage density $P_1(t|r)$, one first
defines the cumulative distribution $C_1(t|r) = \int_{0}^{t}
d\tau \, P_1(\tau|r) = 1- \mathrm{erf}((r-r_{1})/\sqrt{4Dt})$.  Suppose $u$ is
drawn uniformly from the unit interval.  Setting $u=C_1(t_u|r)$ and
solving for $t_u$ gives
\begin{equation}
t_u = \frac{ (r-r_{1})^2}{4D \; \big( \mathrm{ierf}(1-u) \big)^2},
\end{equation}
where $\mathrm{ierf}$ is the inverse error function which can be solved for
numerically in an efficient manner using the secant method.  The set
${t_u}$ are then drawn from the first-passage distribution.

The task of drawing from the distribution $P_2(t|r)$ shown in the
bottom panel of Fig.~\ref{fig:absorption} is readily
accomplished by drawing a random number $p$ uniformly on $(0,1)$ and
then solving the implicit equation $C_2(t_u|r) = p$ for the time $t_u$,
where $C_2(t|r)$ is the cumulative distribution $C_2(t|r) = \int_0^t d\tau
\, P_2(\tau|r)$.  The cumulative distribution can be computed
numerically by applying the Stehfest algorithm to form the inverse
Laplace transform of $\tilde{C}_2(s|r) = \tilde{P}_2 (s|r)/s$ (see
bottom-most panel of Fig.~\ref{fig:absorption}).

\bibliographystyle{apsrev}

\end{document}